# The iCanClean Algorithm: How to Remove Artifacts using Reference Noise Recordings

Ryan J. Downey[1*] and Daniel P. Ferris[1]

[1] J. Crayton Pruitt Family Department of Biomedical Engineering, University of Florida, Gainesville FL, USA

[*]Disclosure: R. J. Downey is the inventor of a pending patent related to the concepts described in the present manuscript.

## Abstract

Data recordings are often corrupted by noise, and it can be difficult to isolate clean data of interest. For example, mobile electroencephalography is commonly corrupted by motion artifact, which limits its use in real-world settings. Here, we describe a novel noise-canceling algorithm that uses canonical correlation analysis to find and remove subspaces of corrupted data recordings that are most strongly correlated with subspaces of reference noise recordings. The algorithm, termed iCanClean, is computationally efficient, which may be useful for real-time applications, such as brain computer interfaces. In future work, we will quantify the algorithm's performance and compare it with alternative cleaning methods.

## 1.Introduction
Noise affects the fidelity of data recordings in many experiments. Algorithms for removing noise could lead to better scientific conclusions from experimental studies. There are many options for removing noise from data recordings, such as frequency-based filtering [1], adaptive filtering [2], wavelets [3], [4], independent component analysis [3], [5], and other blind source separation techniques [3], [4]. However, there is no single algorithm which is optimal for all scenarios. Efficacy of these approaches varies depending on the makeup of the recordings (e.g., transient versus static noise, large versus small amplitude noise, etc.). Here we describe a novel algorithm, termed iCanClean, that may have applications in high-density electroencephalography and other biomedical recording modalities.

## 2. The iCanClean Algorithm
The iCanClean algorithm consists of four steps. First, canonical correlation analysis is used to identify candidate noise components that project onto both the corrupted data recordings as well as the reference noise recordings. Second, a subset of noise components is selected for removal. Third, the projection from the bad components to the data channels is calculated. Fourth, the projected noise components are directly subtracted from the data channels. This process can be applied both to a large window of data and to a smaller sliding window to deal with static and/or transient noise.

Additional mathematical detail is subsequently provided, with key equations written in MATLAB format using built-in MATLAB functions. **Variables** are bolded. *Functions* are italicized. A list of variables is provided in Table 1, along with their description and dimensions.

*Table 1: List of variables*

| Symbol | Description | Dimensions |
|---|---|---|
| **A** | Unmixing matrix (data) | $N_{Data}$ x $N_{Comp}$ |
| **B** | Unmixing matrix (noise) | $N_{Noise}$ x $N_{Comp}$ |
| **BadCompActivity** | Noise components selected for removal | **T** x $N_{Bad}$ |
| **BadCompList** | Indexed list of bad components | $N_{Bad}$ x 1 |
| $N_{Bad}$ | Number of noise components selected for removal | Scalar |
| $N_{Comp}$ | Number of candidate noise components | Scalar |
| $N_{Data}$ | Number of data channels | Scalar |
| $N_{Noise}$ | Number of noise channels | Scalar |
| **ProjectionMatrix** | Relates noise component space to data channel space | $N_{Bad}$ x $N_{Data}$ |
| **ProjectedNoise** | Bad noise components projected onto the data channels | **T** x $N_{Data}$ |
| **R** | Correlation between each ($U_i$ , $V_i$) pair<br>i = [1, 2, …, $N_{Comp}$] | $N_{Comp}$ x 1 |
| **T** | Number of time points (samples) | Scalar |
| **Thresh** | User defined threshold for identifying bad components | Scalar |
| **U** | Candidate noise components (calculated from data channels) | **T** x $N_{Comp}$ |
| **V** | Candidate noise components (calculated from noise channels) | **T** x $N_{Comp}$ |
| **X** | Corrupted data recordings | **T** x $N_{Data}$ |
| $X_{Clean}$ | Cleaned data | **T** x $N_{Data}$ |
| $X_{MC}$ | Mean centered data | **T** x $N_{Data}$ |
| **Y** | Reference noise recordings | **T** x $N_{Noise}$ |
| $Y_{MC}$ | Mean centered noise | **T** x $N_{Noise}$ |

## Definitions

Let **X** be the corrupted data recordings the user wishes to clean with dimensions **T** x $N_{Data}$, where **T** is the number of time points (or samples) and $N_{Data}$ is the number of data channels. Similarly, let **Y** be the reference noise recordings with dimensions **T** x $N_{Noise}$, where $N_{Noise}$ is the number of reference noise channels.

## Step 1

Given corrupted data to clean (**X**) and reference noise recordings (**Y**), canonical correlation analysis is used to identify latent sources of noise (i.e., candidate noise components) in common to both **X** and **Y**.

[ **A** , **B** , **R** , **U** , **V** ] = *canoncorr*( **X** , **Y** );

**A** is an unmixing matrix that converts corrupted data recordings to candidate noise components as **U** = **X**$_{MC}$***A** , where **X**$_{MC}$ is the mean centered data. Similarly, **B** is an unmixing matrix that converts reference noise recordings to a second set of candidate noise components as **V** = **Y**$_{MC}$***B**. Finally, **R** is a vector which contains the correlation between each unique (**U**$_i$, **V**$_i$) pair, where **U**$_i$ and **V**$_i$ are the i$^{th}$ columns of **U** and **V**, respectively. The number of candidate noise components identified, **N**$_{Comp}$ , depends on the rank of the data. Specifically, **N**$_{Comp}$ = *min*( *rank*(**X**) , *rank*(**Y**) ). Therefore, **N**$_{Comp}$ ≤ *min*( **N**$_{Data}$ , **N**$_{Noise}$ ).

With the iCanClean approach, we use corrupted data recordings as one set of inputs to canonical correlation analysis and reference noise recordings as the second set of inputs. Canonical correlation seeks to find the subspaces of these two datasets which are maximally correlated with each other. Because the corrupted data recordings and reference noise recordings both contain noise, canonical correlation analysis will identify hidden noise components that are common to both datasets.

Canonical correlation analysis returns candidate components in ranked order. Thus, the noise component pair with the strongest relationship (i.e., largest **R**$^2$ correlation) appears first (**U**$_1$ , **V**$_1$). The next noise component pair (**U**$_2$ , **V**$_2$) has the second largest **R**$^2$ correlation and is independent of the first component pair, and so forth. The noise components identified by canonical correlation do not depend on how strongly the noise sources project onto the data channels or noise channels. Thus, both large and small noise sources are identified. In a subsequent step, candidate noise components will be appropriately scaled to match the channels being cleaned.

Step 2

A bad subset of components is identified. We suggest a basic thresholding technique where all components with an **R**$^2$ value ≥ a user-defined threshold are rejected. Alternative approaches can be employed as well, for example, by examining the frequency content of the candidate noise components. The user has a choice of whether to select **U** or **V**, or a combination of **U** and **V** as their noise components. The best choice may vary by each specific application. As an example, we assume here that the user wishes to use mixtures (or subspaces) of the data channels to clean the data channels themselves (i.e., use **U** to clean **X**).

**BadCompList** = *find*( **R**.^2 ≥ **Thresh** );

**BadCompActivity** = **U**( : , **BadCompList** );

Step 3

The projection from the noise components onto the channels is calculated. We recommend using a least squares solution (e.g., using matrix division in MATLAB). Alternative options for calculating the projection include applying a Moore-Penrose pseudoinverse to the **A** and/or **B** unmixing matrices, but we have found that it does not perform as well in practice.

**X**$_{MC}$ = **X** - *mean*(**X**);

**ProjectionMatrix** = *mldivide*( **BadCompActivity** , **X**$_{MC}$ );

**ProjectedNoise** = **BadCompActivity** * **ProjectionMatrix**;

Step 4

The projected noise is subtracted from the corrupted data channels.

**X**$_{Clean}$ = **X** - **ProjectedNoise**;

Because the noise components are calculated as linear mixtures of the original recordings (**X** and/or **Y**) and because the projection onto the channels is also linear, iCanClean noise cancellation resembles a spatial filter. This filter can be static or time-varying depending on the application.

## 3. Discussion

In the present work, we described a novel noise canceling algorithm which cleans corrupted data using reference noise recordings. The iCanClean algorithm has potential for cleaning various biomedical signals such as electroencephalography and electromyography. The algorithm is computationally efficient, which may be useful for real-time applications such as brain computer interfaces. In future work, we will examine iCanClean's ability to remove artifacts from electroencephalography using an electrical head phantom, similar to [6], where ground-truth signals are known. This will allow us to quantify its performance relative to alternative cleaning methods.